# Landau-Zener transition in quadratic-nonlinear two-state systems

## A.M. Ishkhanyan

*Institute for Physical Research of NAS of Armenia, 0203 Ashtarak-2, Armenia*

**Abstract.** A comprehensive theory of the Landau-Zener transition in quadratic nonlinear two-state systems is developed. A compact analytic formula involving elementary functions only is derived for the final transition probability. The formula provides a highly accurate approximation for the whole rage of the variation of the Landau-Zener parameter.

The linear-in-time term-crossing two-state model by Landau [1], Zener [2], Stückelberg [3], and Majorana [4] is a long-standing quantum paradigm dating back to the 1930s. It describes the transition dynamics in a two-level quantum-mechanical system with a time-dependent Hamiltonian when the energy separation between the two states is linearly swept in time. This is a well-appreciated approximation that has played a prominent role in the study of a number of important physical phenomena in many branches of contemporary physics ranging from radiation-matter interactions to collision physics. Since the realization of the Bose-Einstein condensates in dilute atomic gases [5], different nonlinear generalizations of this problem have been suggested and explored both theoretically and experimentally. It should be noted that there are two basic modifications of the original Landau-Zener model which differ by the basic type of the involved nonlinearity. In the first case the nonlinearity is accounted for by adding *cubic terms* to the basic *linear* equations describing the original *linear* two-level problem (see, e.g., [6-10]). In contrast, in photo- and magnetoassociation of ultracold atoms an effective two-state problem is encountered where the equations of motion do not involve linear terms so that the basic physical process is principally of a nonlinear nature [11, 12]. In the present article I present a comprehensive analysis of a model described by a *quadratic nonlinear* set of two-level equations that have been a subject of considerable interest during the last years [11-22]. This is a basic version of the nonlinear two-state problem arising in all the nonlinear field theories involving a Hamiltonian with a 2:1 resonance [23]. Applying a two-term ansatz proposed earlier to describe the time-dynamics of the transition probability [22], I construct a highly accurate formula for the final Landau-Zener transition probability. The formula is compact, involves elementary functions only, and is applicable for the whole variation range of the input Landau-Zener parameter.

I consider the following semiclassical nonlinear two-state model treating the atomic and molecular modes as classical fields [11, 12]:

$$i\frac{da_1}{dt} = U(t)e^{-i\delta(t)}\bar{a}_1 a_2,$$
$$i\frac{da_2}{dt} = \frac{U(t)}{2}e^{i\delta(t)}a_1 a_1,$$
(1)

where $t$ is the time, $a_1$ and $a_2$ are the atomic and molecular state probability amplitudes, $\bar{a}_1$ is the complex conjugate of $a_1$, $U(t)$ is the Rabi frequency, and the detuning modulation function $\delta(t)$ is the integral of the detuning $\varepsilon$ ($\varepsilon = d\delta/dt$) of the associating field frequency from that of the transition from the atomic state to the molecular one. This model is encountered, for example, in the theories of cold atom production in atomic Bose-Einstein condensates via laser Raman photoassociation or magnetic Feschbach resonance [24] and in the second harmonic generation in nonlinear optics [25].

For the Landau-Zener model under consideration the Rabi frequency is constant, $U = U_0 = \text{const}$; the detuning is assumed to cross the resonance linearly in time at time point $t = 0$: $d\delta/dt = 2\delta_0 t$; and the conventional Landau-Zener parameter $\lambda$ is defined as $\lambda = U_0^2/\delta_0$. System (1) describes a lossless process so that the total number of particles is conserved: $|a_1|^2 + 2|a_2|^2 = \text{const} = 1$. Finally, I treat the basic case when the system starts from the all-atomic state: $a_1(-\infty) = 1$ and $a_2(-\infty) = 0$.

In order to describe the *time evolution* of the transition probability $p(t) = |a_2(t)|^2$ (note that because of the convention applied here, $p(t) \in [0, 1/2]$), my colleagues and I have previously introduced [22] a two-term ansatz involving three variational parameters, $A$, $C_1$, and $\lambda_1$,

$$p = p_0(A,t) + C_1 \frac{p_{LZ}(\lambda_1, t)}{p_{LZ}(\lambda_1, \infty)},$$
(2)

where, $p_{LZ}(\lambda_1, t)$ is the solution of the *linear* Landau-Zener problem for an effective $\lambda_1$ and $p_0(A,t)$ is a root of the following *quartic* polynomial equation,

$$\frac{\lambda}{4t^2} = \frac{p_0(p_0 - \beta_1)(p_0 - \beta_2)}{9(p_0 - \alpha_1)^2(p_0 - \alpha_2)^2},$$
(3)

where the involved parameters are defined as

$$\alpha_{1,2} = \frac{1}{3} \mp \frac{1}{6}\sqrt{1+\frac{6A}{\lambda}}, \quad \beta_{1,2} = \frac{1}{2} \mp \sqrt{\frac{A}{2\lambda}}. \tag{4}$$

Note that both $p_0(A,t)$ and $p_{LZ}(\lambda_1,t)$ satisfy the initial condition $p(t=-\infty)=0$.

The proposed ansatz is constructed using an exact nonlinear third-order ordinary differential equation obeyed by the transition probability $p(t)$ (Eq. (2), Ref. [22]). I first examine the strong interaction limit of large Landau-Zener parameter, $\lambda \gg 1$, and note that the adiabatic approximation leads to a divergent, at $t \to +\infty$, solution. In order to overcome this divergence, I introduce an augmented (involving a variational parameter $A$) first-order nonlinear equation, the solution of which is the first term $p_0(A,t)$ of ansatz (2). For an appropriately chosen positive $A$, this function accurately describes the time evolution of the system in the time interval covering the prehistory up to the resonance point and an interval after the resonance has been crossed. However, $p_0$ is a monotonically increasing function that does not incorporate the oscillations which start at a certain time point after the resonance has been passed. In order to describe this feature as well, I add a second term, that is, I put $p = p_0 + u$, and examine the exact equation for this correction term $u(t)$. Discussing now the behavior of the system at $t \gg 1$ and neglecting, for a while, the (small) nonlinear terms, we arrive at a linear equation obeyed by a scaled solution to the linear Landau-Zener problem for an effective parameter $\lambda_1$: $u \sim C_1\, p_{LZ}(\lambda_1,t)$. This is the second term of ansatz (2). Finally, via an appropriate choice of $\lambda_1$ and the scaling parameter $C_1$, I account for, to a very good extent, the neglected terms of the equation for $u(t)$ and thus achieve an accurate description of the whole time evolution of the system. The proposed two-term decomposition suggests that in the strong interaction limit the time dynamics of the Landau-Zener transition effectively consists of the essentially nonlinear process of resonance crossing followed by dumped oscillations of a basically linear nature that begin after the resonance has been crossed.

It has been demonstrated numerically that the introduced ansatz produces highly accurate results for any $\lambda$. Namely, the numerical simulations show that for any given value of the input Landau-Zener parameter, $\lambda \in [0,\infty)$, one can always find $A$, $C_1$, and $\lambda_1$ so that function (2) accurately fits the numerical solution to the exact equation for the molecular state probability in the whole time domain – the graphs produced by the formula are practically indistinguishable from the numerical solution. For arbitrary time points, the absolute error is commonly of the order of $10^{-4}$ (a slightly less accurate result, $\sim 10^{-3}$, is observed for points in a small region embracing the first local maximums and minimums of $p(t)$ after the resonance crossing point has been passed) and for the final transition probability $p(+\infty)$ the proposed approximation assures an absolute error of the order of $10^{-5}$ [22].

Further, it has been shown that the variational constant $\lambda_1$, that is, the effective Landau-Zener parameter involved in the linear solution $p_{LZ}(\lambda_1, t)$, is expressed in terms of the final transition probability $p(t = +\infty)$:

$$\lambda_1 = \lambda \, [1 - 3 p(+\infty)]. \qquad (5)$$

Then, to construct analytic approximations for the remaining two variational parameters $A(\lambda)$ and $C_1(\lambda)$, and, hence, to eventually determine the final transition probability written in terms of these parameters as

$$p(+\infty) = \left( \frac{1}{2} - \sqrt{\frac{A}{2\lambda}} \right) + C_1, \qquad (6)$$

a two-parametric fit involving the Gauss hypergeometric function $_2F_1$ [26] has been applied (Eqs. (11) and (12), Ref. [22]).

The principal development I report in the present article is that I show that the scaling parameter $C_1(\lambda)$ can be chosen as

$$C_1 = \frac{P_{LZ}}{4} \sqrt{\frac{2}{\lambda}} A, \qquad (7)$$

where $P_{LZ}$ is the linear Landau-Zener transition probability $P_{LZ} = 1 - e^{-\pi \lambda}$. This particular choice that is justified by examining the next approximation term turns out to be rather productive since it suggests a simpler, *one-parametric* fit. Note that the proposed choice causes only a slight decrease in the accuracy of the fit for the function $p(t)$. A slight increase in the deviation from the numerical solution is observed only for the points of the small region embracing the first local maximums and minimums of the function $p(t)$ after the resonance crossing point has been passed. However, I note that the accuracy of the approximation remains of the same order as before, when the more general two-parametric fit was applied. Importantly, this approach does not affect the accuracy of the approximation for the final transition probability $p(+\infty)$. In the meantime, the proposed choice leads to simple analytic formulas involving elementary functions only for both $A(\lambda)$ and $p(+\infty)$.

The appropriate form of the Gauss hypergeometric function $_2F_1$ to be applied to fit the variational function $A(\lambda)$ if the parameters $\lambda_1(\lambda)$ and $C_1(\lambda)$ are fixed by Eqs. (5) and (7) is given as follows

$$A = \frac{\lambda}{2} \cdot {}_2F_1\left(1, 1+a; 2; -\frac{\lambda^2}{2}\right) \qquad (8)$$

(compare with Eq. (11), Ref. [22]). Note that now the acting variational *constant*, $a$ (this constant should not depend on $\lambda$), is one of the two upper parameters of the hypergeometric function, not the lower one, as applied in Ref. [22]. The immediate advantage of varying the upper parameter of the hypergeometric function of the given form [i.e., with the parameters as indicated in Eq. (8)] is that in this case one derives analytic expressions in terms of elementary functions. Indeed, for an arbitrary $a$ the function $A(\lambda)$ is reduced to

$$A = \frac{1}{a\lambda}\left(1 - \left(1 + \frac{\lambda^2}{2}\right)^{-a}\right). \qquad (9)$$

Accordingly, the final transition probability is written as

$$p(+\infty) = \frac{1}{2} - \frac{a_0(2 - P_{\text{LZ}})}{\lambda}\sqrt{1 - \left(1 + \frac{\lambda^2}{2}\right)^{-a}}, \qquad (10)$$

where $a_0 = 1/\sqrt{8a}$. This formula is the main result of the present article.

Since the limit $\lambda \to +\infty$ holds $P_{\text{LZ}} = 1$ and for a positive $a$ the term under the square root tends to unity, the asymptote of $p(+\infty)$ is given as

$$p(+\infty) = \frac{1}{2} - \frac{a_0}{\lambda}. \qquad (11)$$

This power-law dependence has been indicated by several authors (see, e.g., [14-22]). My colleagues and I have estimated the value of the constant $a_0$ as $a_0 = (2/3)/\pi \approx 0.2122$ [14]. Further studies improved the estimate to $0.2214$ [15]. Given that the *exact* value of this constant is established as $(\ln 2)/\pi$ [21] I finally obtain

$$a_0 = \frac{\ln 2}{\pi} \approx 0.22063560 \quad \text{and} \quad a = \frac{1}{8a_0^2} \approx 2.56778606. \qquad (12)$$

The derived solution defines a fairly good approximation. Comparison of the final transition probability given by the analytic formula (10) with the numerical result is made in Figs. 1(a) and 1(b). It is seen that for the whole variation range of the input Landau-Zener parameter $\lambda$, deviation of the formulas from the numerical result does not exceed $3 \times 10^{-3}$. We note that starting already from $\lambda = 3$ the formula provides five-digit accuracy.

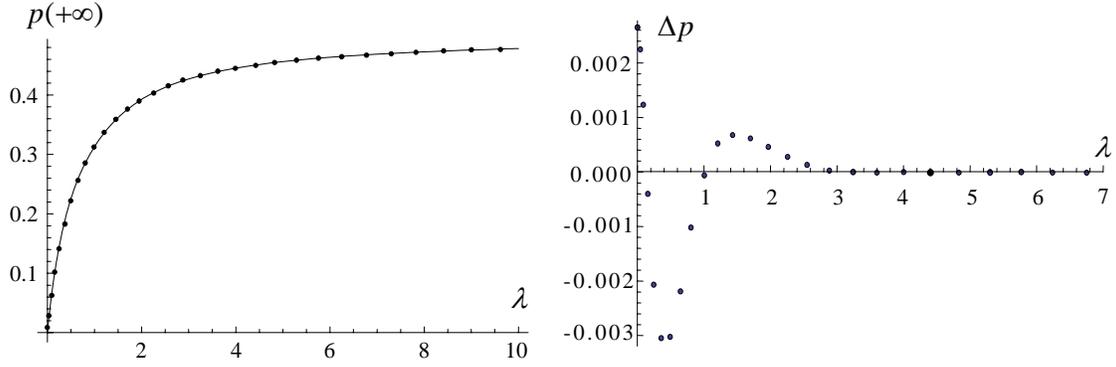

Fig. 1. (a) Final transition probability to the molecular state as a function of $\lambda$. Circles show the numerical result and the solid line is calculated using Eq. (10). (b) Deviation of the approximation defined by Eqs. (10) and (12) from the numerical solution. Starting from $\lambda = 3$ Eq. (10) provides at least five-digit accuracy.

Since all the parameters involved in the general ansatz (2) describing the *whole time dynamics* of the system are eventually written in terms of the principal variational parameter $A$, in order to fulfill the development we examine the behavior of this parameter. The dependence $A(\lambda)$ is shown in Fig. 2. I note that two clearly marked interaction regimes are observed: the weak coupling regime corresponds to $\lambda < 1$, and the strong interaction occurs at $\lambda > 1$ ($A$ reaches its maximum $0.252396$ at $\lambda = 0.947445$). The asymptotic behavior of the parameter $A$ in these two regions is $A \sim \lambda/2$ at $\lambda \to 0$ and $A \sim 1/(a\lambda)$ when $\lambda \to +\infty$.

To summarize, we I developed a comprehensive theory of the Landau-Zener transition in quadratic nonlinear two-state systems describing the ultracold molecule formation dynamics in degenerate quantum gases in the mean-field Gross-Pitaevskii approximation. I have derived a compact analytic formula (10) for the final Landau-Zener transition probability involving elementary functions only that provides a highly accurate approximation for the whole range of the variation of the input Landau-Zener parameter.

This research has been conducted within the scope of the International Associated Laboratory IRMAS. The work was supported by the Armenian National Science and Education Fund (ANSEF Grants No. 2009-PS-1692 and No. 2010-MP-2186).

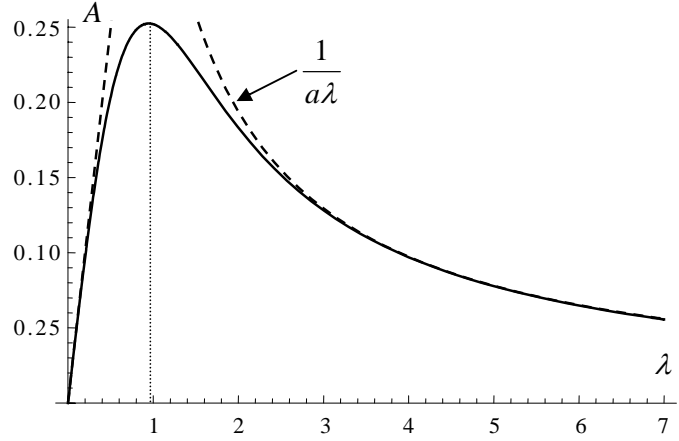

Fig. 2. The principal variational parameter $A$ as a function of the Landau-Zener parameter $\lambda$. The function reaches maximum at $\lambda = 0.947445$. The weak and strong interaction asymptotes are given as $A \sim \lambda/2$ at $\lambda \to 0$ and $A \sim 1/(a\lambda)$ when $\lambda \to +\infty$.